\documentclass[doublecol,a4paper]{epl2} 

\usepackage{footmisc}
\usepackage{amsmath,amssymb,enumerate}
\usepackage{graphicx}

\newcommand{\bee}{\begin{eqnarray}}
\newcommand{\eee}{\end{eqnarray}}
\newcommand{\be}{\begin{equation}}
\newcommand{\ee}{\end{equation}}

\def\eq#1{{Eq.~(\ref{#1})}}

\title{New physics and signal-background interference in associated
  $pp\to HZ$ production}

\abstract{~We re-investigate electroweak signal-background
  interference in associated Higgs production via gluon fusion in the
  presence of new physics in the top-Higgs sector. Considering the
  full final state $pp \to b \bar b \ell^+\ell^-$ ($\ell=e,\mu$), we
  discuss how new physics in the top-Higgs sector that enhances the
  $ZZ$ component can leave footprints in the $HZ$ limit setting. In
  passing we investigate the phenomenology of a class of new physics
  interactions that can be genuinely studied in this process.}
\author{Christoph Englert\inst{1}\and Rogerio Rosenfeld\inst{2}\and
  Michael Spannowsky\inst{3}\and Alberto Tonero\inst{2}}
\shortauthor{C. Englert, R. Rosenfeld, M. Spannowsky, A. Tonero}

\institute{
  \inst{1} SUPA, School of Physics and Astronomy, University of
  Glasgow, Glasgow G12 8QQ, United Kingdom
  
  \inst{2} ICTP-SAIFR \& IFT UNESP, Rua Dr.\ Bento Teobaldo Ferraz
  271, 01140-070, S\~ao Paulo, Brazil

  \inst{3} Institute for Particle Physics Phenomenology, Department
  of Physics, Durham University,\\Durham DH1 3LE, United
  Kingdom
  }

\begin{document}

\maketitle

\section{Introduction}
\label{sec:intro}

After the Higgs discovery in 2012 and initial property
measurements~\cite{Chatrchyan:2012ufa,Aad:2012tfa} in the so-called
$\kappa$ framework, the phenomenology community has now moved towards
understanding constraints in the dimension six effective field theory
(EFT) extension of the Standard Model (SM), which provides a
theoretically clean and well-defined approach to constrain the
presence of new physics interactions with minimal
assumptions~\cite{Buchmuller:1985jz,Hagiwara:1986vm,Giudice:2007fh,Grzadkowski:2010es,Contino:2013kra}.

The field of Standard Model EFT has seen a rapid development
recently. Not only have the run I measurements by ATLAS and CMS been
interpreted in terms of the dimension six EFT
extension~\cite{Azatov:2012bz,Corbett:2012ja,Corbett:2012dm,Espinosa:2012im,Plehn:2012iz,Carmi:2012in,Peskin:2012we,Dumont:2013wma,Djouadi:2013qya,Lopez-Val:2013yba,Englert:2014uua,Ellis:2014dva,Ellis:2014jta,Falkowski:2014tna,Corbett:2015ksa,Buchalla:2015qju,Aad:2015tna,Berthier:2015gja,Englert:2015hrx},
but the EFT framework has also been extended to next-to-leading
order~\cite{Passarino:2012cb,Jenkins:2013zja,Jenkins:2013sda,Jenkins:2013wua,Alonso:2013hga,Hartmann:2015oia,Ghezzi:2015vva,Hartmann:2015aia,Grober:2015cwa}.
Measurement strategies that take into account these corrections via
renormalization group improved calculations have been presented
in~\cite{Isidori:2013cga,Englert:2014cva}.

Due to the large number of effective operators that are relevant to
Higgs physics, it becomes essential to collect information from all
possible processes related to the Higgs boson, especially at the LHC run
II and the future high luminosity phase. Since a single effective
operator can contribute to different processes, there are correlations
among them that can be used to find bounds on the Wilson coefficients
of different operators.  Measurements of associated Higgs
production~\cite{Isidori:2013cga,Biekoetter:2014jwa,Ellis:2014jta},
Higgs+jet
production~\cite{Baur:1989cm,Harlander:2013oja,Banfi:2013yoa,Grojean:2013nya,Schlaffer:2014osa,Buschmann:2014twa,Langenegger:2015lra},
top quark-associated and multi-Higgs \cite{Barr:2013tda,Barr:2014sga,Papaefstathiou:2015paa,Azatov:2015oxa,He:2015spf} production and the recently developed Higgs
off-shell measurements in $gg\to
ZZ$~\cite{Kauer:2012hd,Kauer:2013cga,Englert:2014aca} will be pivotal
to obtain a fine-grained picture of potential compatibility of the
Higgs discovery with the SM expectation.  In particular, the latter
production mechanism has been motivated as an excellent candidate to
constrain new physics effects by exploiting large momentum transfers
to break degeneracies of new physics interactions in the on-shell
Higgs
phenomenology~\cite{Azatov:2014jga,Cacciapaglia:2014rla,Englert:2014ffa,Buschmann:2014sia}.

Similarly, high momentum transfers in associated Higgs production
$pp\to HZ$ are sensitive probes of new
interactions~\cite{Englert:2013vua,Harlander:2013mla,Ellis:2014jta,Hespel:2015zea}.
The reason is the existence of a destructive interference between the
triangle and box contributions in the SM that can be lifted by new or
anomalous couplings. Furthermore, the high momentum transfer provides
another avenue to discriminate the Higgs signal from the background
relying on jet substructure
methods~\cite{Butterworth:2008iy,Soper:2011cr, Soper:2010xk,Altheimer:2013yza,Butterworth:2015bya}.

While jet-substructure analyses provide an extremely versatile and
adaptable tool in new physics and Higgs searches, the mass resolution
of Higgs decays $H\to b\bar b $ in such a search is a limiting
factor. This becomes a challenge especially if cross sections or
beyond the SM-induced deviations thereof become small for large
backgrounds.

\begin{figure}[t!]
\begin{center}
	\includegraphics[width=0.48\textwidth]{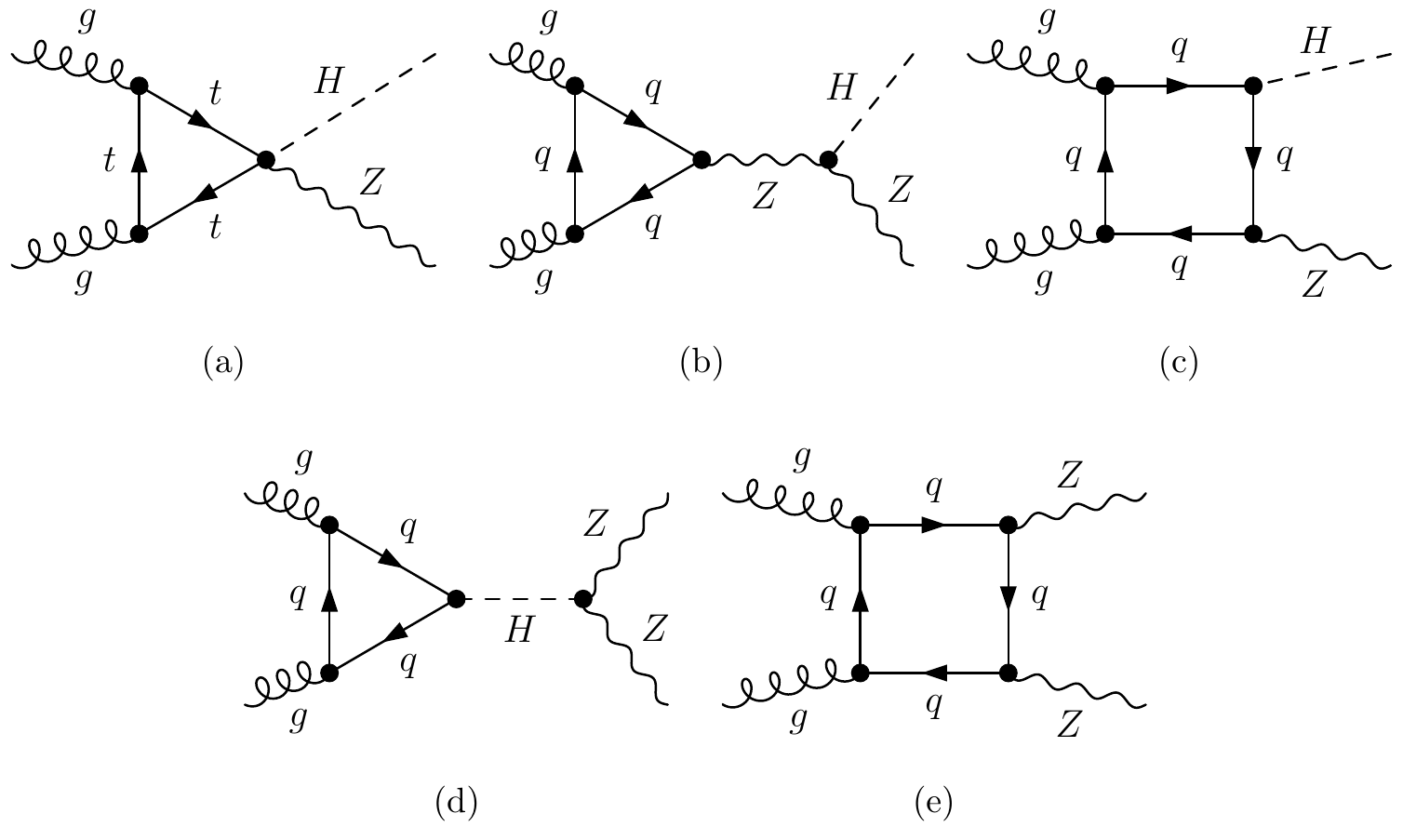}
	\caption{\label{feyn}Representative Feynman diagrams 
  contributing to $pp\to (H,Z)Z\to b\bar b \ell^+\ell^-$; we suppress the Higgs and $Z$ boson decays.}
  \vspace{-0.6cm}
\end{center}
\end{figure}


It is known that gluon fusion-induced associated Higgs
production~\cite{Kniehl:1990iva,Matsuura:1990ba,Altenkamp:2012sx},
while only contributing $\sim 10\%$ of the inclusive $HZ$ production
cross
section~\cite{Hamberg:1990np,Han:1991ia,Ciccolini:2003jy,Brein:2003wg,Ferrera:2011bk,Brein:2011vx,Denner:2011id,Banfi:2012jh,Dawson:2012gs,Brein:2012ne,Goncalves:2015mfa,Ferrera:2014lca,Campbell:2016jau},
becomes relevant at large momentum transfers due to the top quark
threshold~\cite{Englert:2013vua,Harlander:2013mla}.  A similar
argument applies to the non-decoupling of $gg\to H \to ZZ$ at high
momentum transfers~\cite{Glover:1988fe,Kauer:2012hd,Kauer:2013cga}.
Therefore the same type of physics can enhance both $pp \to HZ$ and
$pp \to ZZ$.  We are therefore tempted to ask the following question:
when studying the full final state $pp \to b\bar b \ell^+\ell^-$ as
signal for $pp \to H (\to b\bar b) Z (\to \ell^+\ell)$ \footnote{The
  Higgs decay to leptons, i.e. $pp \to HH \to b\bar b \ell^+\ell^-$ is
  numerically negligible.} for kinematics that allow the discovery of
the Higgs boson in associated production, how important is the
irreducible $pp \to Z (\to b\bar b) Z (\to \ell^+\ell)$ background
keeping in mind an imperfect $H\to b\bar b$ resolution?

To answer this question we organise this letter as follows.  First we
introduce a minimal set of operators which impact the two
contributions $pp\to HZ$ and $pp\to ZZ$ in a different way, but
necessarily related through gauge invariance. We then investigate the
phenomenology of high $p_T$ final states at the parton
level. Subsequently, we show how our findings translate to the fully
hadronized final state before we conclude.

\section{New physics effects in gluon initiated $HZ$ production}
\label{sec:newphys}
Gluon-initiated associated production has been shown to contribute
significantly to $pp\to HZ$ in the boosted regime at the LHC and
important consequences for new physics searches can be obtained by
looking at this
process~\cite{Harlander:2013mla,Englert:2013vua,Goncalves:2015mfa}. New
physics can potentially modify associated Higgs production both in the
quark- and gluon-initiated channels. The quark-initiated channel may
be altered at leading-order through modified Higgs
couplings~\cite{Englert:2014cva} or at next to leading-order through
the influence of new particles or effective operators in
loops~\cite{Harlander:2013mla,Englert:2013tya}. Similarly, the
gluon-initiated channel may receive corrections through modified Higgs
and top couplings to SM states.

In principle, all dimension six operators that are relevant for the
Higgs sector should be considered since at the very least they can
change the Higgs width, which affects the full partonic final
state. However, several of these operators are already constrained
from other observables, such as the $Z$-pole properties measured at LEP1.
In order to keep our discussion transparent, we will focus on only two
operators that are weakly constrained and are relevant for Higgs
production (we adopt the parameterisation
of~\cite{Contino:2013kra,Alloul:2013naa,Bylund:2016phk}):
\begin{alignat}{5}
  \label{eop1} 
  {\cal{O}}_{Ht}=&\frac{i\bar c_{Ht}}{\upsilon^2}
  (\bar t_R \gamma^\mu t_R)(\Phi^\dagger \overleftrightarrow{D}_\mu \Phi)\,,\\
 \label{eop2} 
 {\cal{O}}_t=&
 -\,\frac{\bar c_t}{\upsilon^2} y_t \Phi^\dagger \Phi\, \Phi^\dagger\cdot \bar Q_L \,  t_R  + \text{h.c.}
\end{alignat}
with hermitian covariant derivative $\Phi^\dagger
\overleftrightarrow{D}_\mu \Phi= \Phi^\dagger (D_\mu \Phi)-
(D_\mu\Phi)^\dagger\Phi$, and $\Phi$ being the weak doublet that
contains the physical Higgs $\Phi \supset H$.
The operator in~\eq{eop1} modifies the coupling of the right-handed
top quark to the $Z$ boson $\bar t_R t_R Z$ by a factor proportional
to the $\bar c_{Ht}$ coefficient
\begin{equation} 
  \label{eq:modferm}
  \frac{2}{3}g\frac{s_W^2}{c_W}\to \frac{2}{3}g\frac{ s_W^2}{c_W} +g\frac{\bar c_{Ht}}{2 c_W} \,.
\end{equation}
It affects the $Ztt$ coupling but not $Htt$ and introduces a new
$ttHZ$ coupling.  As required by gauge invariance, the derivative
coupling of the top quark to the neutral Goldstone boson gets also
shifted by the same quantity.  Couplings to left-handed quark doublets
are constrained by data on $Z\to b\bar b$ and will not change the
qualitative outcome of our discussion.\footnote{Interactions of this
  type can typically arise in composite Higgs
  scenarios~\cite{Ferretti:2014qta}, which will also leave footprints
  in $q\bar q \to HZ$ as a function of the fine-tuning parameter
  $v^2/f^2$, where $f$ is the pion decay constant analogue.}
Operators of this form but involving light fermions are constrained by
precision electroweak measurements $|c_{Hu}|\lesssim 2\%$ and assuming
a trivial flavor structure of the UV dynamics will directly constrain
the interaction of Eq.~\eqref{eop1}, which is otherwise unconstrained
at the tree level by electroweak precision data and has no impact on Higgs
decays (see, e.g.~\cite{Contino:2013kra} for a comprehensive
discussion). Higher order corrections, however, re-induce a
dependence, see \cite{Dror:2015nkp}. We will ignore this potential
constraint for the time being, but will come back to it later.

\begin{figure}[t!]
\begin{center}
  \includegraphics[height=0.27\textwidth]{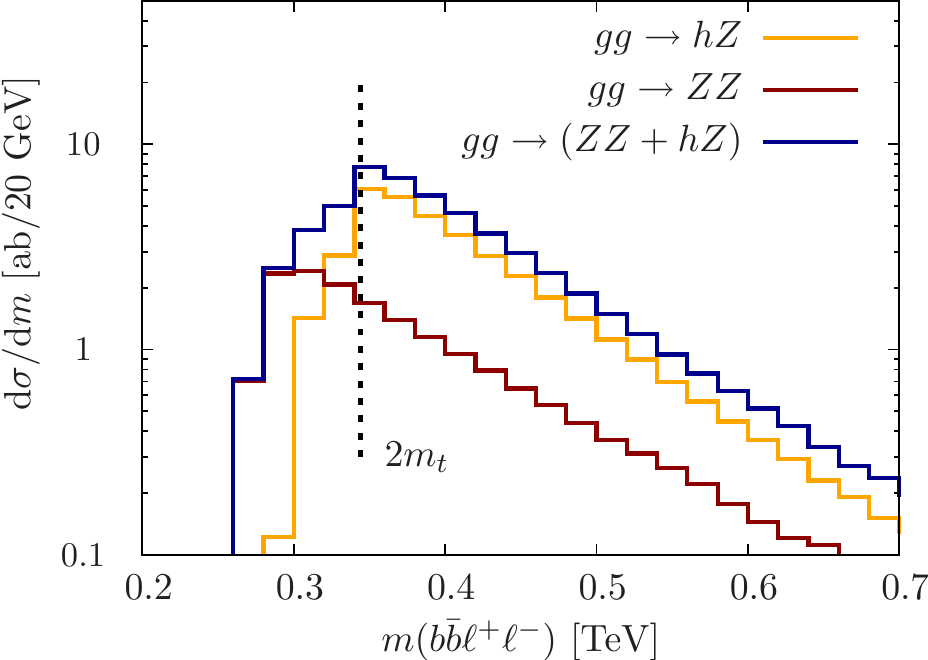}
  \caption{Invariant mass distribution of the $b \bar b \ell^+ \ell^-$
    (in this plot $\ell=\mu$) system for the final state $g g \to
    b\bar b \ell^+ \ell^-$ in the SM and the phase space region
    $p_T(\ell^+\ell^-)\gtrsim 100~\text{GeV}$ relevant for a boosted
    $H\to b \bar b $ analysis.  \label{minv}}
    \vspace{-0.5cm}
\end{center}
\end{figure}

The operator in \eq{eop2} modifies the top Yukawa coupling by a factor
proportional to Wilson coefficient $\bar c_{t}$,
%
$  y_t \to y_t (1+\bar c_t )$,
while leaving the top mass as in the SM with a simple redefinition of
the top quark field.  The non-derivative couplings of the top quark to
the neutral Goldstone boson are unchanged.

We show in Fig.~\ref{feyn} the relevant Feynman diagrams for $pp \to
HZ$ and $pp \to ZZ$ ignoring the diagrams involving the unphysical
Goldstone bosons.  Note in particular the new effective vertex $\bar t t H
Z $ introduced by operator \eq{eop1}, not present in the SM, which
gives rise to the Feynman diagram contribution to the gluon-initiated
amplitude shown in Fig.~\ref{feyn} (a), and which may affect the
cancellation between triangle and box diagrams for $pp \to HZ$ in the
SM, leading to an enhanced cross section.  This cancellation is also
impacted by the change in the top Yukawa coupling introduced by operator
\eq{eop2}.  In fact, the effect of a flipped top Yukawa coupling
(i.e., with a coupling of opposite sign with respect to the SM,
corresponding to $\bar c_t = -2$) on $pp \to HZ$ was studied in
\cite{Hespel:2015zea}.

Together these operators provide a parameterisation that allow us to
``template'' the $gg\to ZZ$ and $gg\to HZ$ components of the full
partonic final state $pp\to b \bar b l^+l^-$ in a gauge invariant
fashion, and therefore gives us a well-defined approach to study the
signal-background interference in this final state. Note that since
these operators only modify the $ttH$ and $ttZ$ couplings, they do not
affect the tree-level $q \bar q \to HZ$ process.  Only the operator
\eq{eop2} changes the Higgs branching ratios (by a few percent in the
relevant $BR(H \to b \bar b)$ in the cases explored here) and it has
been taken into account.

The new interactions arising from \eq{eop1} and \eq{eop2} were
implemented using {\sc{FeynRules}}~\cite{Alloul:2013bka}.  We
calculate the one-loop gluon-initiated $gg \to (HZ+ZZ) \to b \bar b
\ell^+ \ell^-$ production amplitudes using the {\sc{FeynArts}},
{\sc{FormCalc}} and {\sc{LoopTools}}~\cite{Hahn:1998yk,Hahn:2000kx}
framework which we interface with {\sc{Vbfnlo}}~\cite{Arnold:2008rz}
to perform the phase space integration and generate events in the Les
Houches standard and keep the full quark mass dependencies throughout.
We pass these events to
{\sc{Herwig++}}~\cite{Bahr:2008pv} for showering and
hadronization. The $q\bar q$-initiated process is simulated with {\sc
  MadGraph5}~\cite{Alwall:2014hca} using an identical input parameter
setting and passed through {\sc{Herwig++}} to obtain the full hadronic
final state. The respective samples are normalised to the NLO QCD
predictions of the SM~\cite{Hamberg:1990np,Altenkamp:2012sx}.  We use
a $K-$factor of 1.2 and 1.8 for $qq$ and $gg$-initiated processes
respectively. We focus on collisions at 13 TeV centre of mass energy.

\subsection{Parton level analysis}
\label{sec:planalysis}
Before we analyse the full hadron level, it is worthwhile to
re-investigate the order of magnitude of expected interference effects
between the $gg\to HZ$ and $gg\to ZZ$ parts in the full $p p \to HZ +
ZZ$ final state (see also~\cite{Goncalves:2015mfa} for an earlier
discussion). To this end, we show in Fig.~\ref{minv} the parton level
comparison of the invariant mass distribution between $HZ$ and $ZZ$
production for gluon-initiated $b \bar b \ell^+\ell^-$ (in this case
$\ell=\mu$) production.  Note the rise of the cross section near the
$2 m_t$ threshold. For these selection requirements we find a
SM cross section of 0.9~fb (including the flat $K-$factor). 
A choice of $\bar{c}_{Ht}=1,\bar{c}_t=0$ increases this cross section by 70\%. A quantitatively
 identical enhancement can be achieved for $\bar{c}_{Ht}=0,\bar{c}_t\simeq 0.33$.

Signal-background interference between the two contributions is in
general a small effect and the relative size of $HZ$ dominates over
$ZZ$ as a consequence of the relative branching ratio suppression of
$H\to b\bar b$ (60\%) and $Z\to \bar b b$ (15\%). This is left
unchanged for changes in $\bar c_t$~\cite{Goncalves:2015mfa}, however,
there will be modifications from Eq.~\eqref{eop2}.

\begin{figure}[t!]
\begin{center}
  \includegraphics[height=0.27\textwidth]{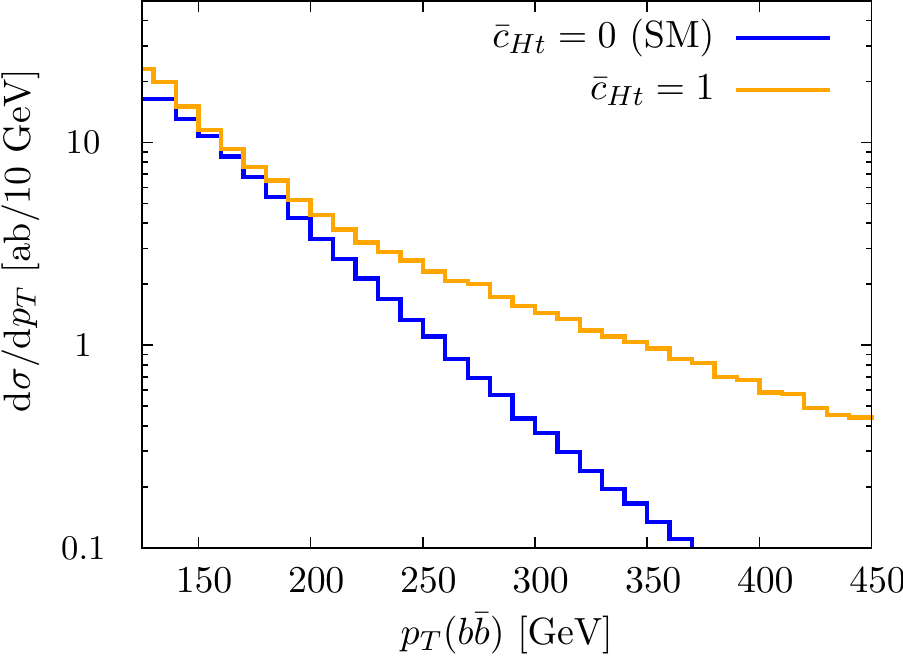}
  \caption{Transverse Higgs $p_T$ distribution and its sensitivity to
    the operator $\bar c_{Ht}$. It can be seen that the boosted regime
    $p_T(\ell^+\ell^-)\simeq p_T({b\bar b})\gtrsim 150~\text{GeV}$ is
    highly sensitive to the operator Eq.~\eqref{eop1} which also
    modifies the continuum $ZZ$ production.  \label{pt}}
     \vspace{-0.5cm}
\end{center}
\end{figure}

In order to obtain a first estimate of the sensitivity to the
effective operators, we consider first the process $pp \to (HZ+ZZ) \to
b \bar b \ell^+ \ell^-$ again at parton level. Based on the event
simulation described above, we select events with
\begin{equation}
	 p_T(\ell^+\ell^-)>150~\text{GeV}\,,~110~\text{GeV} < m(b\bar b)<140~\text{GeV}
\end{equation}
As an example, we show in Fig.~\ref{pt} the effect of $\bar c_{Ht} =
1$.  One can see that this operator can dramatically impact the
boosted Higgs regime due to the lifting of the SM cancellation and
also the derivative nature of the induced
coupling~\cite{Bylund:2016phk}.

\begin{figure}[t!]
\begin{center}
\includegraphics[width=0.42\textwidth]{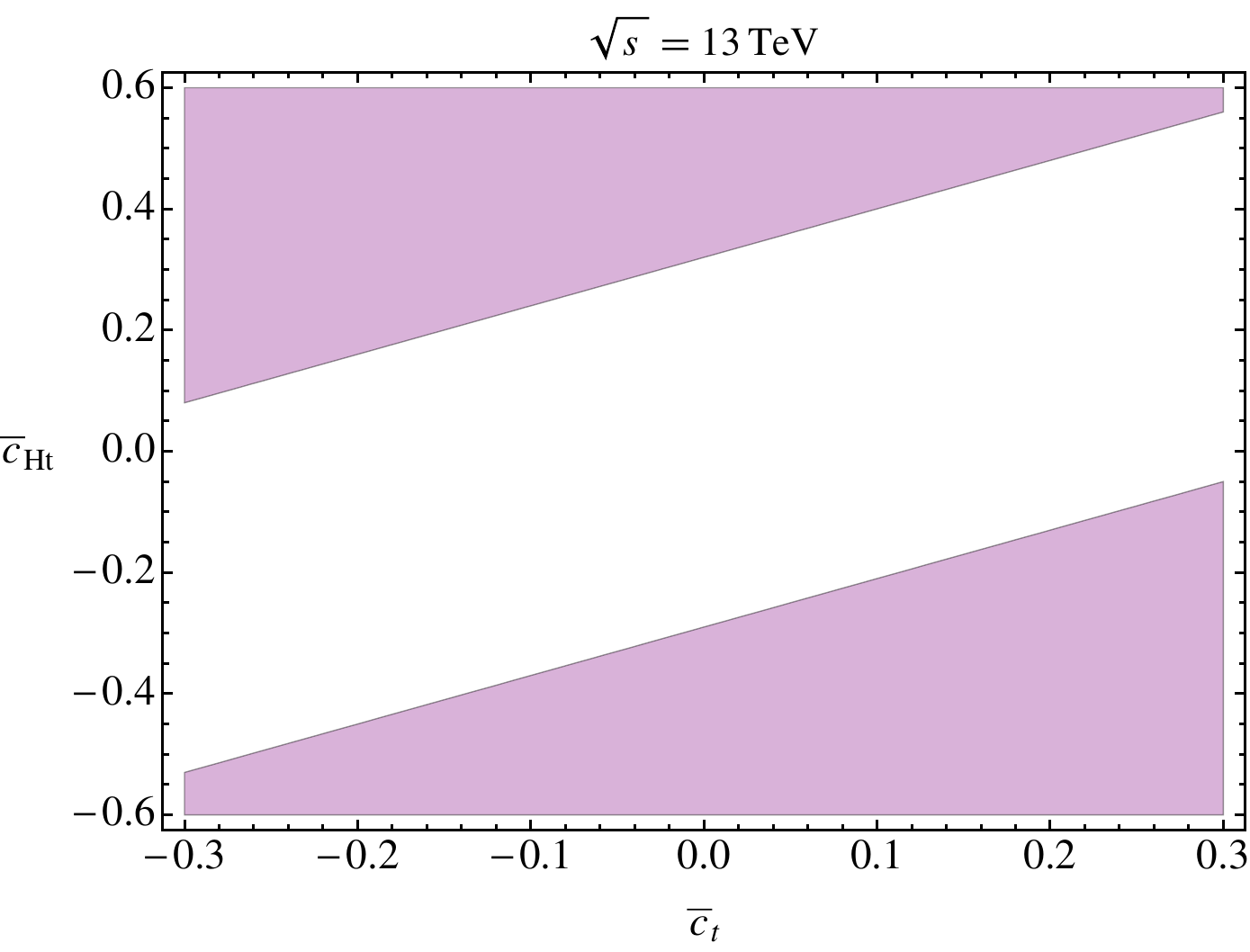}
\caption{\label{bounds} Projected sensitivity of the boosted parton
  level analysis of $pp \to b\bar b \ell^+\ell^-$ in the conventions
  of Eqs.~\eqref{eop1} and \eqref{eop2}, the shaded region is excluded
  at 95\% confidence level for the ideal parton level setting
  described in the text, for ${\cal{L}}=100~\text{fb}^{-1}$.}
  \vspace{-0.5cm}
\end{center}
\end{figure}

In order to derive exclusion regions in the $(\bar c_t, \bar c_{Ht})$
plane we perform a log-likelihood hypothesis test based on a shape
comparison of the $p_T(b\bar b)$ distribution using the CL$_s$ method~\cite{Junk:1999kv,James:2000et,Read:2002hq}. 

In Fig.~\ref{bounds} we show the expected exclusion for a luminosity
of 100 fb$^{-1}$ based on our parton level results. While the resonant
and continuum $ZZ$ contributions are largely suppressed, the
gauge-invariant extension of the top loop-induced $gg\to ZZ$
diagram\footnote{One can understand the modification of the $Zt\bar t$
  interaction as replacing $H\to \langle H \rangle$.} introduces the
$t\bar t HZ$ interaction. The result of Fig.~\ref{bounds} indicates
that the modification according to operator Eq.~\eqref{eop1}, even for
small choices in agreement with precision
analyses~\cite{Contino:2013kra} can in principle impact the limit
setting procedure in associated Higgs production through sculpting the
$p_T(b\bar b)$ distribution, especially when marginalising over
Eq.~\eqref{eop2} in a global fit where degenerate operator directions
will influence the expected exclusion.

One might worry about the validity of the Effective Field
  Theory in our analysis. This issue has been a subject of recent
  discussion, see e.g. \cite{ Englert:2014cva, Contino:2016jqw}. The
  coefficients of the dimension-6 operators can be related to the
  scale $M$ where new physics appears by $\bar{c} \approx g^2
  v^2/M^2$, where $g$ is a coupling constant of the heavy states with
  SM particles. Further suppression factors arise in the case in which
  an operator is generated at loop level. We can therefore put an upper bound in the new mass scale from requiring that the underlying theory is strongly coupled, i.e., $g = 4 \pi$: $M< 4 \pi v/\sqrt{c} \approx 3$ TeV for $c = {\cal O}(1)$. Since our analysis relies on $p_T<1$ TeV we do not violate this upper bound.

\subsection{Showering and hadronization}
\label{sec:hadalysis}

The results of the parton analysis detailed in the previous section
are known to change substantially when we turn to the full hadron
level final state and perform a realistic
reconstruction~\cite{Englert:2013vua}. Based on the event generation
strategy outlined above, we apply typical $HZ$ final state selection
cuts by
\begin{enumerate}[(i)]
\item requiring exactly 2 oppositely charged same-flavor leptons
  satisfying $|\eta_\ell|<2.5$, $p_T(\ell)>30~\text{GeV}$,
\item require that these leptons are compatible with the $Z$ boson mass:
  $80~\text{GeV} < m(\ell^+\ell^-)<100~\text{GeV}$,
\item and require boosted topologies $p_T(\ell^+\ell^-)>200$ GeV.
\item We then perform a typical BDRS
  analysis~\cite{Butterworth:2008iy}: All remaining hadronic activity
  is clustered using {\sc{FastJet}}~\cite{Cacciari:2011ma} into a
  Cambridge-Aachen fat jet with $R=1.2$. The boosted Higgs candidate
  jet has to satisfy $p_{T,j}>200~\text{GeV}$ and at least one such
  object is required in $|\eta|<2.5$. The fat jet is filtered,
  mass-dropped and double b-tagged with a b-tag efficiency of 60\% (2\%
  fake rate), yielding a total efficiency of 36\%.
\item Higgs candidates are required to be compatible with
  $110~\text{GeV} < m(b\bar b)<140~\text{GeV}$ evaluated on the
  b-tagged subjets.
\end{enumerate}
While the high-$p_T$ selection is enough to remove the biggest
background $t\bar t$ almost entirely, jet-substructure approaches
remove the QCD-induced $b\bar b$ production modes from the selection
to a large extent, leaving $Z+$jet production as a dominant background
(or calibration tool). The Higgs mass resolution quoted in (v) is a
key factor in the boosted analysis to allow signal vs. background
extraction in the first place (and veto SM $q\bar q$-induced $ZZ$
production). However as mentioned before the gluon-induced $ZZ$
contribution could in principle be enhanced through the operator
discussed previously, thus adding more significantly to the region (v)
than expected in the SM and at parton level due to shower and
hadronization effects.

\begin{figure}[b!]
\begin{center}
\includegraphics[width=0.42\textwidth]{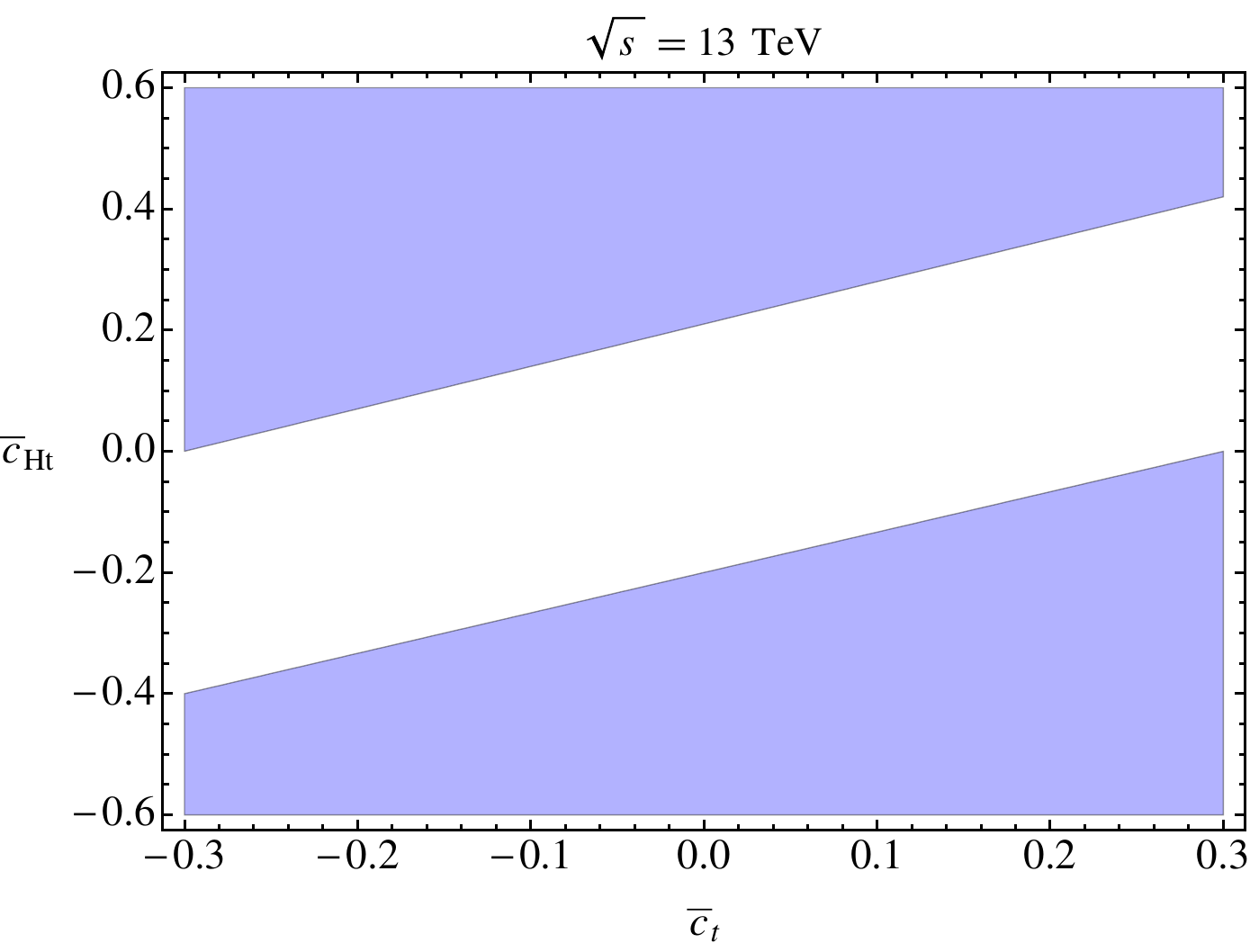}
\caption{\label{bounds2} Projected exclusion at 95\% CL$_s$ (blue shaded region) 
	of the boosted hadron level analysis of $pp \to b\bar b \ell^+\ell^-$ 
	at $3~\text{ab}^{-1}$ integrated luminosity.}
\end{center}
\end{figure}

After these analysis steps one typically obtains a cross section of
$\sim 0.2$ fb for the SM which includes both $q\bar q$ and
$gg$-initiated processes. And again we find the impact of $HZ$ far
more dominant than $ZZ$. As expected, the lowered statistical yield
when taking into account the full reconstruction efficiencies requires
a larger luminosity to set limits. Setting limits, we obtain a result
comparable to the parton analysis of the previous section for
$3~\text{ab}^{-1}$, see Fig.~\ref{bounds2}. This means that when
including the constraints from complementary Higgs measurements at
this luminosity, which are expected to limit $|\bar{c}_t|\lesssim
10^{-2}$~\cite{Englert:2015hrx}, the presence of $\bar c_{Ht}$ for
trivial flavor structures, {\it i.e.} at the level of $\bar c_{Ht} =
\bar c_{Hu}$ is difficult to constrain and can practically be
neglected when working with this assumption. However, associated Higgs
production provides test of non-trivial beyond the SM flavour
structures, which can be combined with direct $t\bar t Z$ searches
(see
e.g.~\cite{Rontsch:2014cca,Khachatryan:2014ewa,Aad:2015eua,Buckley:2015lku,Tonero:2014jea,Dror:2015nkp,Bylund:2016phk}). Comparing to the projections of \cite{Rontsch:2014cca}, $-0.13<\bar c_{Ht} < 0.64$, we see that associated Higgs production can be expected to provide a additional discriminating power to complementary $t\bar t Z$ searches. It should be noted that our results do not reflect systematic uncertainties from both theoretical and experimental sources and are therefore very likely to worsen, in particular in a global fit when more operators are included. In particular, the theoretical uncertainties due to missing higher orders in $gg\to HZ$ are currently large for boosted kinematics $\sim{\cal{O}}(30\%)$ \cite{Altenkamp:2012sx}. Potential improvements in particular related to experimental systematics are hard to foresee at this stage in the LHC programme, but our results suggest that boosted Higgs analysis should continue to receive attention.

\section{Summary and Conclusions}
\label{sec:sum}
In this letter we have re-investigated electroweak signal-background
interference in gluon-initiated associate Higgs production in light of
expected efficiencies and selection requirements of the fully
hadronized final state. While $HZ+ZZ$ signal-background interference
is suppressed, new physics effects that impact $pp\to ZZ$ can also
leave footprints in boosted analyses $pp\to HZ$ through new
interactions related by gauge invariance. However, a robust limit
setting in this channel will require a large luminosity.  Even at
these large luminosities the constraints on $\bar c_{Ht}$ will not be
competitive with electroweak precision constraints under the
assumption of a trivial flavor structure (as commonly done in Higgs
fits at this stage in the LHC phenomenology programme). Relaxing this
assumption, associated Higgs production via gluon fusion can act as a
test of this hypothesis, especially when other measurements point
towards the SM.

\subsection{Acknowledgements}
CE thanks the organisers of the 2014 ICTP-SAIFR GOAL Workshop, where
this work was initiated and Marco Farina for helpful discussions. MS
is supported by the European Commission through ITN
PITN-GA-2012-316704 (``HiggsTools''). AT is supported by the S\~ao
Paulo Research Foundation (FAPESP) under grants 2011/11973-4 and
2013/02404-1. RR is partially supported by the FAPESP grant
2011/11973-4 and by a CNPq research grant.  RR thanks C\'edric
Delaunay and Heidi Rzehak for early discussions on some topics of this
paper.


\bibliographystyle{eplbib}
\bibliography{paper}

\begin{thebibliography}{100}
\expandafter\ifx\csname url\endcsname\relax\def\url#1{\texttt{#1}}\fi

\bibitem{Chatrchyan:2012ufa}
\Name{Chatrchyan S. \etal} \REVIEW{Phys.Lett.B}{716}{2012}{30}.

\bibitem{Aad:2012tfa}
\Name{Aad G. \etal} \REVIEW{Phys.Lett.B}{716}{2012}{1}.

\bibitem{Buchmuller:1985jz}
\Name{Buchmuller W. \and Wyler D.} \REVIEW{Nucl. Phys.B}{268}{1986}{621}.

\bibitem{Hagiwara:1986vm}
\Name{Hagiwara K., Peccei R.~D., Zeppenfeld D. \and Hikasa K.} \REVIEW{Nucl.
  Phys.B}{282}{1987}{253}.

\bibitem{Giudice:2007fh}
\Name{Giudice G.~F., Grojean C., Pomarol A. \and Rattazzi R.}
  \REVIEW{JHEP}{06}{2007}{045}.

\bibitem{Grzadkowski:2010es}
\Name{Grzadkowski B., Iskrzynski M., Misiak M. \and Rosiek J.}
  \REVIEW{JHEP}{10}{2010}{085}.

\bibitem{Contino:2013kra}
\Name{Contino R., Ghezzi M., Grojean C., M{\"u}hlleitner M. \and Spira M.}
  \REVIEW{JHEP}{07}{2013}{035}.

\bibitem{Azatov:2012bz}
\Name{Azatov A., Contino R. \and Galloway J.} \REVIEW{JHEP}{04}{2012}{127}
  [Erratum: JHEP04,140(2013)].

\bibitem{Corbett:2012ja}
\Name{Corbett T., Eboli O. J.~P., Gonzalez-Fraile J. \and Gonzalez-Garcia
  M.~C.} \REVIEW{Phys. Rev.D}{87}{2013}{015022}.

\bibitem{Corbett:2012dm}
\Name{Corbett T., Eboli O. J.~P., Gonzalez-Fraile J. \and Gonzalez-Garcia
  M.~C.} \REVIEW{Phys. Rev.D}{86}{2012}{075013}.

\bibitem{Espinosa:2012im}
\Name{Espinosa J.~R., Grojean C., M{\"u}hlleitner M. \and Trott M.}
  \REVIEW{JHEP}{12}{2012}{045}.

\bibitem{Plehn:2012iz}
\Name{Plehn T. \and Rauch M.} \REVIEW{Europhys. Lett.}{100}{2012}{11002}.

\bibitem{Carmi:2012in}
\Name{Carmi D., Falkowski A., Kuflik E., Volansky T. \and Zupan J.}
  \REVIEW{JHEP}{10}{2012}{196}.

\bibitem{Peskin:2012we}
\Name{Peskin M.~E.} \REVIEW{}{}{2012}{}.

\bibitem{Dumont:2013wma}
\Name{Dumont B., Fichet S. \and von Gersdorff G.} \REVIEW{JHEP}{07}{2013}{065}.

\bibitem{Djouadi:2013qya}
\Name{Djouadi A. \and Moreau G.} \REVIEW{Eur. Phys. J.C}{73}{2013}{2512}.

\bibitem{Lopez-Val:2013yba}
\Name{Lopez-Val D., Plehn T. \and Rauch M.} \REVIEW{JHEP}{10}{2013}{134}.

\bibitem{Englert:2014uua}
\Name{Englert C., Freitas A., M{\"u}hlleitner M.~M., Plehn T., Rauch M., Spira
  M. \and Walz K.} \REVIEW{J. Phys.G}{41}{2014}{113001}.

\bibitem{Ellis:2014dva}
\Name{Ellis J., Sanz V. \and You T.} \REVIEW{JHEP}{07}{2014}{036}.

\bibitem{Ellis:2014jta}
\Name{Ellis J., Sanz V. \and You T.} \REVIEW{JHEP}{1503}{2015}{157}.

\bibitem{Falkowski:2014tna}
\Name{Falkowski A. \and Riva F.} \REVIEW{JHEP}{02}{2015}{039}.

\bibitem{Corbett:2015ksa}
\Name{Corbett T., Eboli O. J.~P., Goncalves D., Gonzalez-Fraile J., Plehn T.
  \and Rauch M.} \REVIEW{JHEP}{08}{2015}{156}.

\bibitem{Buchalla:2015qju}
\Name{Buchalla G., Cata O., Celis A. \and Krause C.} \REVIEW{}{}{2015}{}.

\bibitem{Aad:2015tna}
\Name{Aad G. \etal} \REVIEW{}{}{2015}{}.

\bibitem{Berthier:2015gja}
\Name{Berthier L. \and Trott M.} \REVIEW{JHEP}{02}{2016}{069}.

\bibitem{Englert:2015hrx}
\Name{Englert C., Kogler R., Schulz H. \and Spannowsky M.} \REVIEW{}{}{2015}{}.

\bibitem{Passarino:2012cb}
\Name{Passarino G.} \REVIEW{Nucl. Phys.B}{868}{2013}{416}.

\bibitem{Jenkins:2013zja}
\Name{Jenkins E.~E., Manohar A.~V. \and Trott M.} \REVIEW{JHEP}{10}{2013}{087}.

\bibitem{Jenkins:2013sda}
\Name{Jenkins E.~E., Manohar A.~V. \and Trott M.} \REVIEW{Phys.
  Lett.B}{726}{2013}{697}.

\bibitem{Jenkins:2013wua}
\Name{Jenkins E.~E., Manohar A.~V. \and Trott M.} \REVIEW{JHEP}{01}{2014}{035}.

\bibitem{Alonso:2013hga}
\Name{Alonso R., Jenkins E.~E., Manohar A.~V. \and Trott M.}
  \REVIEW{JHEP}{04}{2014}{159}.

\bibitem{Hartmann:2015oia}
\Name{Hartmann C. \and Trott M.} \REVIEW{JHEP}{07}{2015}{151}.

\bibitem{Ghezzi:2015vva}
\Name{Ghezzi M., Gomez-Ambrosio R., Passarino G. \and Uccirati S.}
  \REVIEW{JHEP}{07}{2015}{175}.

\bibitem{Hartmann:2015aia}
\Name{Hartmann C. \and Trott M.} \REVIEW{Phys. Rev. Lett.}{115}{2015}{191801}.

\bibitem{Grober:2015cwa}
\Name{Gr{\"o}ber R., M{\"u}hlleitner M., Spira M. \and Streicher J.}
  \REVIEW{JHEP}{09}{2015}{092}.

\bibitem{Isidori:2013cga}
\Name{Isidori G. \and Trott M.} \REVIEW{JHEP}{02}{2014}{082}.

\bibitem{Englert:2014cva}
\Name{Englert C. \and Spannowsky M.} \REVIEW{Phys. Lett.B}{740}{2015}{8}.

\bibitem{Biekoetter:2014jwa}
\Name{Biek{\"o}tter A., Knochel A., Kr{\"a}mer M., Liu D. \and Riva F.}
  \REVIEW{Phys. Rev.D}{91}{2015}{055029}.

\bibitem{Baur:1989cm}
\Name{Baur U. \and Glover E. W.~N.} \REVIEW{Nucl. Phys.B}{339}{1990}{38}.

\bibitem{Harlander:2013oja}
\Name{Harlander R.~V. \and Neumann T.} \REVIEW{Phys. Rev.D}{88}{2013}{074015}.

\bibitem{Banfi:2013yoa}
\Name{Banfi A., Martin A. \and Sanz V.} \REVIEW{JHEP}{08}{2014}{053}.

\bibitem{Grojean:2013nya}
\Name{Grojean C., Salvioni E., Schlaffer M. \and Weiler A.}
  \REVIEW{JHEP}{05}{2014}{022}.

\bibitem{Schlaffer:2014osa}
\Name{Schlaffer M., Spannowsky M., Takeuchi M., Weiler A. \and Wymant C.}
  \REVIEW{Eur. Phys. J.C}{74}{2014}{3120}.

\bibitem{Buschmann:2014twa}
\Name{Buschmann M., Englert C., Goncalves D., Plehn T. \and Spannowsky M.}
  \REVIEW{Phys. Rev.D}{90}{2014}{013010}.

\bibitem{Langenegger:2015lra}
\Name{Langenegger U., Spira M. \and Strebel I.} \REVIEW{}{}{2015}{}.

\bibitem{Barr:2013tda}
\Name{Barr A.~J., Dolan M.~J., Englert C. \and Spannowsky M.} \REVIEW{Phys.
  Lett.B}{728}{2014}{308}.

\bibitem{Barr:2014sga}
\Name{Barr A.~J., Dolan M.~J., Englert C., Ferreira~de Lima D.~E. \and
  Spannowsky M.} \REVIEW{JHEP}{02}{2015}{016}.

\bibitem{Papaefstathiou:2015paa}
\Name{Papaefstathiou A. \and Sakurai K.} \REVIEW{JHEP}{02}{2016}{006}.

\bibitem{Azatov:2015oxa}
\Name{Azatov A., Contino R., Panico G. \and Son M.} \REVIEW{Phys.
  Rev.D}{92}{2015}{035001}.

\bibitem{He:2015spf}
\Name{He H.-J., Ren J. \and Yao W.} \REVIEW{Phys. Rev.D}{93}{2016}{015003}.

\bibitem{Kauer:2012hd}
\Name{Kauer N. \and Passarino G.} \REVIEW{JHEP}{08}{2012}{116}.

\bibitem{Kauer:2013cga}
\Name{Kauer N.} \REVIEW{Mod. Phys. Lett.A}{28}{2013}{1330015}.

\bibitem{Englert:2014aca}
\Name{Englert C. \and Spannowsky M.} \REVIEW{Phys. Rev.D}{90}{2014}{053003}.

\bibitem{Azatov:2014jga}
\Name{Azatov A., Grojean C., Paul A. \and Salvioni E.} \REVIEW{Zh. Eksp. Teor.
  Fiz.}{147}{2015}{410} [J. Exp. Theor. Phys.120,354(2015)].

\bibitem{Cacciapaglia:2014rla}
\Name{Cacciapaglia G., Deandrea A., Drieu La~Rochelle G. \and Flament J.-B.}
  \REVIEW{Phys. Rev. Lett.}{113}{2014}{201802}.

\bibitem{Englert:2014ffa}
\Name{Englert C., Soreq Y. \and Spannowsky M.} \REVIEW{JHEP}{05}{2015}{145}.

\bibitem{Buschmann:2014sia}
\Name{Buschmann M., Goncalves D., Kuttimalai S., Schonherr M., Krauss F. \and
  Plehn T.} \REVIEW{JHEP}{02}{2015}{038}.

\bibitem{Englert:2013vua}
\Name{Englert C., McCullough M. \and Spannowsky M.} \REVIEW{Phys.
  Rev.D}{89}{2014}{013013}.

\bibitem{Harlander:2013mla}
\Name{Harlander R.~V., Liebler S. \and Zirke T.} \REVIEW{JHEP}{02}{2014}{023}.

\bibitem{Hespel:2015zea}
\Name{Hespel B., Maltoni F. \and Vryonidou E.} \REVIEW{JHEP}{06}{2015}{065}.

\bibitem{Butterworth:2008iy}
\Name{Butterworth J.~M., Davison A.~R., Rubin M. \and Salam G.~P.}
  \REVIEW{Phys. Rev. Lett.}{100}{2008}{242001}.

\bibitem{Soper:2011cr}
\Name{Soper D.~E. \and Spannowsky M.} \REVIEW{Phys. Rev.D}{84}{2011}{074002}.

\bibitem{Soper:2010xk}
\Name{Soper D.~E. \and Spannowsky M.} \REVIEW{JHEP}{08}{2010}{029}.

\bibitem{Altheimer:2013yza}
\Name{Altheimer A. \etal} \REVIEW{Eur. Phys. J.C}{74}{2014}{2792}.

\bibitem{Butterworth:2015bya}
\Name{Butterworth J.~M., Ochoa I. \and Scanlon T.} \REVIEW{Eur. Phys.
  J.C}{75}{2015}{366}.

\bibitem{Kniehl:1990iva}
\Name{Kniehl B.~A.} \REVIEW{Phys. Rev.D}{42}{1990}{2253}.

\bibitem{Matsuura:1990ba}
\Name{Matsuura T., Hamberg R. \and van Neerven W.~L.} \REVIEW{Nucl.
  Phys.B}{345}{1990}{331}.

\bibitem{Altenkamp:2012sx}
\Name{Altenkamp L., Dittmaier S., Harlander R.~V., Rzehak H. \and Zirke T.
  J.~E.} \REVIEW{JHEP}{02}{2013}{078}.

\bibitem{Hamberg:1990np}
\Name{Hamberg R., van Neerven W.~L. \and Matsuura T.} \REVIEW{Nucl.
  Phys.B}{359}{1991}{343} [Erratum: Nucl. Phys.B644,403(2002)].

\bibitem{Han:1991ia}
\Name{Han T. \and Willenbrock S.} \REVIEW{Phys. Lett.B}{273}{1991}{167}.

\bibitem{Ciccolini:2003jy}
\Name{Ciccolini M.~L., Dittmaier S. \and Kramer M.} \REVIEW{Phys.
  Rev.D}{68}{2003}{073003}.

\bibitem{Brein:2003wg}
\Name{Brein O., Djouadi A. \and Harlander R.} \REVIEW{Phys.
  Lett.B}{579}{2004}{149}.

\bibitem{Ferrera:2011bk}
\Name{Ferrera G., Grazzini M. \and Tramontano F.} \REVIEW{Phys. Rev.
  Lett.}{107}{2011}{152003}.

\bibitem{Brein:2011vx}
\Name{Brein O., Harlander R., Wiesemann M. \and Zirke T.} \REVIEW{Eur. Phys.
  J.C}{72}{2012}{1868}.

\bibitem{Denner:2011id}
\Name{Denner A., Dittmaier S., Kallweit S. \and Muck A.}
  \REVIEW{JHEP}{03}{2012}{075}.

\bibitem{Banfi:2012jh}
\Name{Banfi A. \and Cancino J.} \REVIEW{Phys. Lett.B}{718}{2012}{499}.

\bibitem{Dawson:2012gs}
\Name{Dawson S., Han T., Lai W.~K., Leibovich A.~K. \and Lewis I.}
  \REVIEW{Phys. Rev.D}{86}{2012}{074007}.

\bibitem{Brein:2012ne}
\Name{Brein O., Harlander R.~V. \and Zirke T. J.~E.} \REVIEW{Comput. Phys.
  Commun.}{184}{2013}{998}.

\bibitem{Goncalves:2015mfa}
\Name{Goncalves D., Krauss F., Kuttimalai S. \and Maierh{\"o}fer P.}
  \REVIEW{Phys. Rev.D}{92}{2015}{073006}.

\bibitem{Ferrera:2014lca}
\Name{Ferrera G., Grazzini M. \and Tramontano F.} \REVIEW{Phys.
  Lett.B}{740}{2015}{51}.

\bibitem{Campbell:2016jau}
\Name{Campbell J.~M., Ellis R.~K. \and Williams C.} \REVIEW{}{}{2016}{}.

\bibitem{Glover:1988fe}
\Name{Glover E. W.~N. \and van~der Bij J.~J.} \REVIEW{Phys.
  Lett.B}{219}{1989}{488}.

\bibitem{Englert:2013tya}
\Name{Englert C. \and McCullough M.} \REVIEW{JHEP}{07}{2013}{168}.

\bibitem{Alloul:2013naa}
\Name{Alloul A., Fuks B. \and Sanz V.} \REVIEW{JHEP}{04}{2014}{110}.

\bibitem{Bylund:2016phk}
\Name{Bylund O.~B., Maltoni F., Tsinikos I., Vryonidou E. \and Zhang C.}
  \REVIEW{}{}{2016}{}.

\bibitem{Ferretti:2014qta}
\Name{Ferretti G.} \REVIEW{JHEP}{06}{2014}{142}.

\bibitem{Dror:2015nkp}
\Name{Dror J.~A., Farina M., Salvioni E. \and Serra J.}
  \REVIEW{JHEP}{01}{2016}{071}.

\bibitem{Alloul:2013bka}
\Name{Alloul A., Christensen N.~D., Degrande C., Duhr C. \and Fuks B.}
  \REVIEW{Comput. Phys. Commun.}{185}{2014}{2250}.

\bibitem{Hahn:1998yk}
\Name{Hahn T. \and Perez-Victoria M.} \REVIEW{Comput. Phys.
  Commun.}{118}{1999}{153}.

\bibitem{Hahn:2000kx}
\Name{Hahn T.} \REVIEW{Comput. Phys. Commun.}{140}{2001}{418}.

\bibitem{Arnold:2008rz}
\Name{Arnold K. \etal} \REVIEW{Comput. Phys. Commun.}{180}{2009}{1661}.

\bibitem{Bahr:2008pv}
\Name{Bahr M. \etal} \REVIEW{Eur. Phys. J.C}{58}{2008}{639}.

\bibitem{Alwall:2014hca}
\Name{Alwall J., Frederix R., Frixione S., Hirschi V., Maltoni F., Mattelaer
  O., Shao H.~S., Stelzer T., Torrielli P. \and Zaro M.}
  \REVIEW{JHEP}{07}{2014}{079}.

\bibitem{Junk:1999kv}
\Name{Junk T.} \REVIEW{Nucl. Instrum. Meth.A}{434}{1999}{435}.

\bibitem{James:2000et}
\Name{James F., Perrin Y. \and Lyons L.} (Editors) \Book{{Workshop on
  confidence limits, CERN, Geneva, Switzerland, 17-18 Jan 2000: Proceedings}}
  2000.
\newline\url{http://weblib.cern.ch/abstract?CERN-2000-005}

\bibitem{Read:2002hq}
\Name{Read A.~L.} \REVIEW{J. Phys.G}{28}{2002}{2693} [,11(2002)].

\bibitem{Contino:2016jqw}
\Name{Contino R., Falkowski A., Goertz F., Grojean C. \and Riva F.}
  \REVIEW{}{}{2016}{}.

\bibitem{Cacciari:2011ma}
\Name{Cacciari M., Salam G.~P. \and Soyez G.} \REVIEW{Eur. Phys.
  J.C}{72}{2012}{1896}.

\bibitem{Rontsch:2014cca}
\Name{R{\"o}ntsch R. \and Schulze M.} \REVIEW{JHEP}{07}{2014}{091} [Erratum:
  JHEP09,132(2015)].

\bibitem{Khachatryan:2014ewa}
\Name{Khachatryan V. \etal} \REVIEW{Eur. Phys. J.C}{74}{2014}{3060}.

\bibitem{Aad:2015eua}
\Name{Aad G. \etal} \REVIEW{JHEP}{11}{2015}{172}.

\bibitem{Buckley:2015lku}
\Name{Buckley A., Englert C., Ferrando J., Miller D.~J., Moore L., Russell M.
  \and White C.~D.} \REVIEW{}{}{2015}{}.

\bibitem{Tonero:2014jea}
\Name{Tonero A. \and Rosenfeld R.} \REVIEW{Phys. Rev.D}{90}{2014}{017701}.

\end{thebibliography}

\end{document}